\documentclass{article}
\usepackage[preprint]{neurips_2020}

\usepackage[utf8]{inputenc} 
\usepackage[T1]{fontenc}    
\usepackage{hyperref}       
\usepackage{url}            
\usepackage{booktabs}       
\usepackage{amsfonts}       
\usepackage{nicefrac}       
\usepackage{microtype}      
\usepackage{amsmath}
\usepackage{graphicx}
\usepackage{subfigure}
\usepackage[ruled,longend]{algorithm2e}
\usepackage{multirow}
\usepackage{wrapfig}
\usepackage{mathtools}
\newcommand{\vect}[1]{\boldsymbol{#1}}

\title{Skill Discovery of Coordination in Multi-agent Reinforcement Learning}
\author{
    Shuncheng He\\
    Tsinghua University\\
    Beijing, China\\
    \texttt{hesc16@mails.tsinghua.edu.cn}
    \And
    Jianzhun Shao \\
    Tsinghua University\\
    Beijing, China\\
    \texttt{sjz18@mails.tsinghua.edu.cn} \\
    \AND
    Xiangyang Ji\\
    Tsinghua University\\
    Beijing, China\\
    \texttt{xyji@tsinghua.edu.cn} \\
}
\begin{document}

\maketitle

\begin{abstract}
Unsupervised skill discovery drives intelligent agents to explore the unknown environment without task-specific reward signal, and the agents acquire various skills which may be useful when the agents adapt to new tasks. In this paper, we propose "Multi-agent Skill Discovery"(MASD), a method for discovering skills for coordination patterns of multiple agents. The proposed method aims to maximize the mutual information between a latent code $Z$ representing skills and the combination of the states of all agents. Meanwhile it suppresses the empowerment of $Z$ on the state of any single agent by adversarial training. In another word, it sets an information bottleneck to avoid empowerment degeneracy. First we show the emergence of various skills on the level of coordination in a general particle multi-agent environment. Second, we reveal that the "bottleneck" prevents skills from collapsing to a single agent and enhances the diversity of learned skills. Finally, we show the pretrained policies have better performance on supervised RL tasks.
\end{abstract}
\section{Introduction}
Unsupervised reinforcement learning (RL) allows intelligent agents to learn various skills simultaneously without any extrinsic rewards related to specific tasks \citep{gupta2018unsupervised, eysenbach2018diversity}. Most of unsupervised RL methods utilizes a latent-conditioned policy to optimize an information theoretical objective. The condition of the policy can be associated with a "goal", which is generated randomly, by a prior, or in a heuristic way for exploring new states in the environment. This approach helps the agent to quickly adapt to the tasks requiring the agent to reach some goal states. Also the condition can be perceived as a latent code of high-level skills or options. The agent is driven to learn distinct skills or options which are discriminable from their states, or trajectories \citep{gregor2016variational, achiam2018variational}. These papers show the skills learned without supervision help the agent tackle with challenging tasks with sparse reward, form a option set for hierarchical RL, and provide a good initialization for further training.\par
Ideally, these unsupervised skill discovery algorithms can be seamlessly transplanted to multi-agent reinforcement learning (MARL) environments. However, three problems immediately emerge. First, the nature of MARL emphasizes the interaction and coordination amongst the agents and it is clearly out of consideration of the skills trained by individual agents. How can we train the agents to autonomously focus on the skill of coordination, or their interaction patterns? Second, under the framework of centralized training and decentralized execution due to partial observability, the policies will inevitably converges to suboptimal points, whether with task-specific reward or with unsupervised surrogate reward \citep{mahajan2019maven}. Finally, rather than the environments used in single agent unsupervised RL, the multi-agent environment is highly unstable and volatile from the view of an individual agent. How can the agents retain a discriminable skill?\par
In this paper, we propose a novel algorithm, called \textit{multi-agent skill discovery} (MASD), to address skill discovery on the level of coordination amongst multiple agents. Two key ideas are involved to design MASD. We first introduce a latent variable shared by all agents and maximize the mutual information between the latent variable and the whole set of states. Then we set an "information bottleneck" on individual states, namely, minimizing the mutual information between the latent and states of any single agents in an adversarial way, which forces the policies to learn skills on a higher level of coordination and interaction. Within the scope of implementation, we adopt MADDPG, an actor-critic structured algorithm with centralized training and decentralized execution, to optimize the surrogate objective derived from the two principles stated above.\par
Our work makes three contributions. First, we propose a method for learning skills in multi-agent environments without supervision. Second, we show the empowerment degeneracy and the collapse onto a single agent without the information bottleneck, both on simple demonstrations and particle multi-agent environments. Third, we demonstrate that MASD can learn a series of distinguishable skills of coordination and show initializing with good skills can outperform baseline algorithm on a complex supervised task.
\section{Background}
\subsection{Preliminaries}
Partially observable Markov decision process (POMDP) is an appropriate model to conclude many multi-agent Markov games. POMDPs are formally defined as a tuple $G=\langle S,U,P,X,r,o,\gamma,N\rangle$. $S$ is the set of all possible states in the environment. At each time step $t$, the $i$th agent receives its own observation $x_{t}^{(i)}\in X$. The observation is generated from the internal state $s_{t}\in S$ through a observing lens $x_{t}^{(i)}=o(s_{t},i)$. With a certain policy, the agent chooses action $u_{t}^{(i)}$ from the action space $U$ and send it to the environment. The environment returns a new state $s_{t+1}$ according to state transition probability distribution $P(s_{t+1}|s_{t},\vect{u}_{t})$, where the tuple of actions from all agents is denoted as $\vect{u}_{t}=(u_{t}^{(1)},\dots,u_{t}^{(N)})$, and generates a scalar reward $R_{t}=r(s_{t},\vect{u}_{t})$. After $T$ steps, the episode terminates. In supervised and decentralized scenarios, the agents improve their policies $\pi^{i}(u^{(i)}|x^{(i)})$ to maximize their collective expected accumulative discounted reward $\mathbb{E}_{s_{0},\pi,P}[\sum_{t=0}^{T}\gamma^{t}R_{t}]$.
\subsection{Mutual information and variational inference}
The mutual information $I(S;Z)$ is a general measure of dependency between two random variables. It is defined as the \textit{Kullback-Leibler divergence} between the joint distribution $p(s,z)$ and the product of two distributions $p(s)\cdot p(z)$ as
\begin{equation}
    I(S;Z)=\int_{Z}\int_{S} p(s,z)\log\frac{p(s,z)}{p(s)p(z)}dsdz.
\end{equation}
An alternative expression $I(S;Z)=H(S)-H(S|Z)$ implies when the mutual information is high, the uncertainty of variable $S$ is largely reduced given $Z$. Therefore the mutual information can be interpreted as the empowerment from one variable to another.
In unsupervised RL, a latent random variable $Z$ is introduced as a condition of policies $\pi(a|s,z)$. We hope the latent variable can shed its controllability on successive states, or trajectories. It is straightforward to set the mutual information between $Z$ and the states $S$, as the unsupervised objective. If $I(S;Z)$ is maximized, the behaviour of the agent will change consistently given different values of the latent code.\par
However, estimating and optimizing $I(S;Z)$ can be very challenging. By symmetry we have $I(S;Z)=H(Z)-H(Z|S)$. When the prior of $Z$ is fixed, maximizing $I(S;Z)$ is equivalent to maximizing the negative entropy $-H(Z|S)$. Nevertheless, the posterior distribution $p(z|s)$ is remained unknown, and we cannot compute it directly due to intractability of marginal distribution $p(s)$. Fortunately taking the tool of variational inference, we have a variational lower bound of the objective \citep{blei2017variational}\par
\begin{equation}
    \label{ELBO}
    -H(Z|S)=\mathbb{E}_{s\sim p(s),z\sim p(z)}[\log p(z|s)]\geq\mathbb{E}_{s\sim p(s),z\sim p(z)}[\log q_{\phi}(z|s)].
\end{equation}
In \eqref{ELBO}, $q_{\phi}(z|s)$ is an approximator of the true posterior parameterized with $\phi$. Actually the gap of this inequality is the KL divergence between $p(z|s)$ and $q_{\phi}(z|s)$, which means that the more precise the approximator is, the tighter the lower bound is.\par
In conclusion, we use $r_{z}(s,a)=\log q_{\phi}(z|s)$ as pseudo reward to train the agent. Meanwhile we train the approximator (we call it the discriminator below) with $(z,s)$ pairs stored in a replay memory.
\subsection{Adversarial learning}
In previous papers like DIAYN, VIC, the agent and the discriminator evolve together in a cooperative way \citep{eysenbach2018diversity}. However, in our work, we expect the agents to minimize some kind of mutual information (stated in the next section), which leads the agents and the discriminators to learn adversarially. A precedent work, Generative Adversarial Imitation Learning (GAIL) demonstrates the feasibility of implementing adversarial training in RL \citep{ho2016generative, song2018multi}. It allows two entities to optimize a mini-max objective in which case the two opposite entities co-evolve.
\section{Skill Discovery of Coordination}
In this section, we propose a new method called \textit{multi-agent skill discovery} (MASD). This algorithm dedicates to autonomous discovery of skills of coordinated agents. It indicates that the acquired skills are not affiliated to any single agent, but reflect different patterns in their coordinating behaviour, which is crucial to the agents in cooperative MARL tasks \citep{lowe2017multi}.\par
Inspired by other unsupervised skill discovery methods in single agent RL, the straightforward way is letting the policy conditioned on a sampled latent variable $z$ shared by all agents in each episode, and maximizing the mutual information between the latent $Z$ and the overall state $S$. However it raises a tricky issue: unlike single agent configuration, the overall state of the multi-agent environment is unknown to us in most cases. What we have is a set of observations retrieved from the agents distributed in a map. Therefore, we can only use the combination of all observations, denoted as $\vect{X}=(X^{(1)},\dots,X^{(N)})$. In practice, $\vect{X}$ contains redundant information, i.e., if the $i$th agent and the $j$th agent are mutually visible, both of the observation vectors will include the information of the other agent. To this point, we extract the features necessary for learning through $f(X^{(i)})$, from the full-length observation vector. For convenience, we call $f(X^{(i)})$ the "state" of the $i$th agent, although it is not the actual state of the environment.\par
In summary, the collective objective of all agent is to maximize $I(f(\vect{X});Z)$ where $Z\sim p(z)$ is interpreted as skills. With slight abuse of notation, we use $f(\vect{X})=(f(X^{(1)}),\dots,f(X^{(N)}))$ to denote the set of extracted features. On one hand, the sampled skill controls the set of states visited by multiple agents. On the other hand, the agents make the latent skill distinguishable from the states. The mutual information measures the obedience of the agents to the instruction $Z$. Nevertheless, this dose not automatically imply that the latent variable controls the coordination patterns amongst the agents. There are possibilities that maximizing $I(f(\vect{X});Z)$ may result in degeneracy, which means the latent $Z$ solely controls the state of a single agent. This is partially due to the suboptimality trap of decentralized MARL \citep{mahajan2019maven}. Furthermore in a toy experiment, we demonstrate maximizing $I(f(\vect{X});Z)$ can lead to multiple optimal policies but some of them are degenerated.\par
\subsection{Enforcing the policies out of degeneracy}
Intuitively, we are not desired to see the latent $Z$ is clearly discriminable from a single agent. Thus the latent $Z$ is forced to cast its controllability on the relations of the agents. To this end, we reduce every $I(f(X^{(i)});Z)$ and the objective becomes
\begin{equation}
\label{OBJ}
\begin{split}
    \mathcal{F}(\vect{\theta})&=I(f(\vect{X});Z)-\frac{1}{N}\sum_{i=1}^{N}I(f(X^{(i)});Z)\\
    &=-H(Z|f(\vect{X}))+\frac{1}{N}\sum_{i=1}^{N}H(Z|f(X^{(i)}))\\
    &=\mathcal{F}_{1}(\vect{\theta})+\mathcal{F}_{2}(\vect{\theta}).
\end{split}
\end{equation}
Here $\vect{\theta}=(\theta_{1},\dots,\theta_{N})$ contains policy parameters of $N$ agents. The first term suggests it should be accurate to infer skill $Z$ from the combination of all states and the second term guarantees the opaqueness of $Z$ from individual agents. $\mathcal{F}_{1}(\vect{\theta})$ has a variational lower bound by \eqref{ELBO} using a parameterized global discriminator
\begin{equation}
    \mathcal{F}_{1}(\vect{\theta})\geq\mathbb{E}_{s_{0},\pi_{\theta},P}\log q_{\phi}(z|\vect{x})\coloneqq\mathcal{G}_{1}(\vect{\theta},\phi)
\end{equation}
However $\mathcal{F}_{2}(\vect{\theta})$ does not have a non-trivial lower bound. Despite using $N$ local discriminators $\vect{\phi}=(\phi_{1},\dots,\phi_{N})$, what we yield
\begin{equation}
    \mathcal{F}_{2}(\vect{\theta})\leq-\frac{1}{N}\sum_{i=1}^{N}\mathbb{E}_{s_{0},\pi_{\theta},P}\log q_{\phi_{i}}(z|f(x^{(i)}))\coloneqq\mathcal{G}_{2}(\vect{\theta},\vect{\phi})
\end{equation}
is an upper bound of the objective. We can also maximize the entropy $\mathcal{F}_{2}(\vect{\theta})$ in an adversarial way, a resemblance to GAN or GAIL \citep{ho2016generative, goodfellow2014generative}. The mini-max objective becomes
\begin{equation}
    \min_{\phi_{1},\dots,\phi_{N}}\max_{\theta_{1},\dots,\theta_{2}}\mathcal{G}_{2}(\vect{\theta},\vect{\phi})
\end{equation}
Therefore we feed the multi-agent policies with pseudo reward
\begin{equation}
\label{REW}
    r_{z}=\log q_{\phi}(z|f(\vect{x}))-\frac{1}{N}\sum_{i=1}^{N}\log q_{\phi_{i}}(z|f(x^{(i)})).
\end{equation}
Meanwhile we reduce the entropy of the global discriminator $q_{\phi}$ and $N$ local discriminators $q_{\phi_{i}},i=1,\dots,N$ with rollout data $(z,\vect{x})$. Local discriminators endeavor to distinguish latent skill code $z$ from their own states $f(x^{(i)})$ whilst the agents maintain high entropy of the posterior $p(z|f(x^{(i)}))$ to hide the latent from local states. Hence as the agents learn to perform various skills, the entropy regularizer $\mathcal{F}_{2}(\theta)$ prevents the skills from degeneracy on the behaviour of a single agent. 
\subsection{Implementation}
\begin{figure}[htbp]
    \centering
    \includegraphics[width=0.8\textwidth]{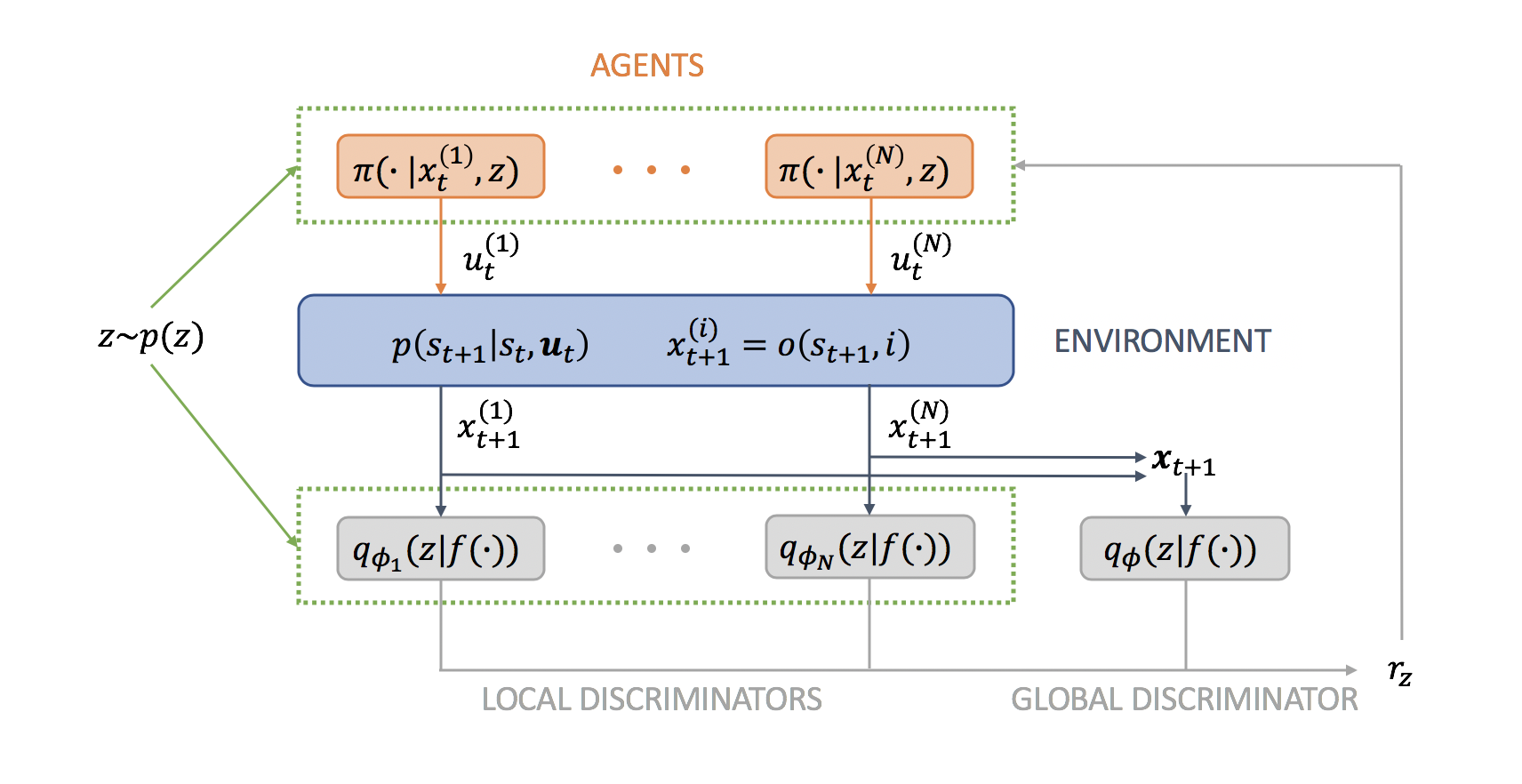}
    \caption{\textbf{MASD} framework. Agents receive pseudo reward computed by discriminators to improve their coordinating skills.}
    \label{framework}
\end{figure}
\textit{Multi-agent deep deterministic policy gradient} (MADDPG) is an actor-critic MARL algorithm, composed of $N$ actors with policy $\pi_{\theta_{i}}(u|x)$ and $N$ critics $Q_{\psi_{i}}(\vect{x},\vect{u})$. MADDPG avoids high variance of classical policy gradient methods and alleviates the difficulties brought by non-stationarity in multi-agent Q-learning. Therefore we choose MADDPG as the basic learning framework to optimize our proposed objective \eqref{OBJ}. At the same time, we train the global discriminator and $N$ local discriminators with supervision loss. The overall structure is depicted in Fig. \ref{framework}. Notice that the latent space can either be continuous or categorical. When $z$ is sampled from a $k$-category uniform distribution, the discriminator is equipped with categorical cross entropy loss. And when $z$ is sampled from a uniform distribution $U[-1,1]$, the discriminator is optimized by $L_{1}$ or $L_{2}$ loss. Choosing $L_{1}$ or $L_{2}$ loss depends on our hypothesis on the distribution family of the posterior $p(z|x)$. The $L_{1}$ loss corresponds to Laplacian distribution and $L_{2}$ loss is related to Gaussian distribution. Disregarding what the latent space is, we denote the loss of the global discriminator as $\mathcal{L}_{\phi}(z)$ and the losses of local discriminators as $\mathcal{L}_{\phi_{i}}(z)$. Our method is summarized in Algorithm \ref{ALGO}.\par
\begin{algorithm}[H]
\label{ALGO}
Initialize parameters $\phi$ and $\phi_{i},\theta_{i},\psi_{i}$ for each $i=1,\dots,N$\\
Initialize replay memory $\mathcal{D}_{\text{rl}}\leftarrow\{\}$ and $\mathcal{D}_{\text{disc}}\leftarrow\{\}$\\
\For{\text{each episode}}{
    Sample a skill $z\sim p(z)$\\
    Get observations $\vect{x}_{0}$ for all agents\\
    \For{\text{each time step} t}{
        $u_{t}^{(i)}\sim\pi_{\theta_{i}}(u|x)$\\
        Apply $\vect{u}_{t}$ to the environment and get observations $\vect{x}_{t+1}$\\
        Generate pseudo reward $r_{z}$ by \eqref{REW}\\
        $\mathcal{D}_{\text{rl}}\leftarrow\mathcal{D}_{\text{rl}}\cup\{(\vect{x}_{t},\vect{u}_{t},z,r_{z},\vect{x}_{t+1})\}$\\
        $\mathcal{D}_{\text{disc}}\leftarrow\mathcal{D}_{\text{disc}}\cup\{(f(\vect{x}_{t+1}),z)\}$
    }
    \For{\text{each update step}}{
        Sample a minibatch $\mathcal{B}_{\text{rl}}$ from $\mathcal{D}_{\text{rl}}$\\
        Compute MADDPG actor loss and critic loss with $\mathcal{B}_{\text{rl}}$\\
        Update $\theta_{i}$ and $\psi_{i}$ for each actor and critic\\
        Sample a minibatch $\mathcal{B}_{\text{disc}}$ from $\mathcal{D}_{\text{disc}}$\\
        Compute $\mathcal{L}_{\phi}(z)$ and $\mathcal{L}_{\phi_{i}}(z)$ for each agent with $\mathcal{B}_{\text{disc}}$\\
        Update $\phi$ and $\phi_{i}$ for the global discriminator and each local discriminator
    }
}
\caption{MASD}
\end{algorithm}
In practice we use different variants of pseudo reward. First we multiply the second term in \eqref{REW} by $\beta$ to balance the global discriminability and the local opaqueness. Second, we can replace the mean $\frac{1}{N}\sum_{i=1}^{N}\log q_{\phi_{i}}(z|f(x^{(i)})$ by the minimum $\min_{i}\log q_{\phi_{i}}(z|f(x^{(i)})$ to emphasize the worst case across all agents.
\begin{table}
  \caption{Comparison of 2 optimal solutions, mutual information in bits}
  \label{EXAM}
  \centering
  \begin{tabular}{cccccc}
    \toprule
    Solution    & $Z$       & $(X'^{(1)},X'^{(2)})$ & $I(\vect{X}',Z)$ & $I(X'^{(1)};Z)$ & $I(X'^{(2)};Z)$\\
    \midrule
    \multirow{2}{*}{A}   & 0   & (0,1),(1,0) &\multirow{2}{*}{1} &\multirow{2}{*}{0} &\multirow{2}{*}{0} \\\cmidrule{2-3}
                    & 1   & (1,1),(0,0) &\\\midrule
    \multirow{2}{*}{B}   & 0   & (0,0),(0,1) &\multirow{2}{*}{1} &\multirow{2}{*}{1} &\multirow{2}{*}{0}\\\cmidrule{2-3}
                    & 1   & (1,0),(1,1) &\\
    \bottomrule
  \end{tabular}
\end{table}
\section{Experiments}
\subsection{Empowerment degeneracy: a toy example}

\begin{wrapfigure}{r}{0.6\textwidth}
\begin{center}
    \includegraphics[width=0.6\textwidth]{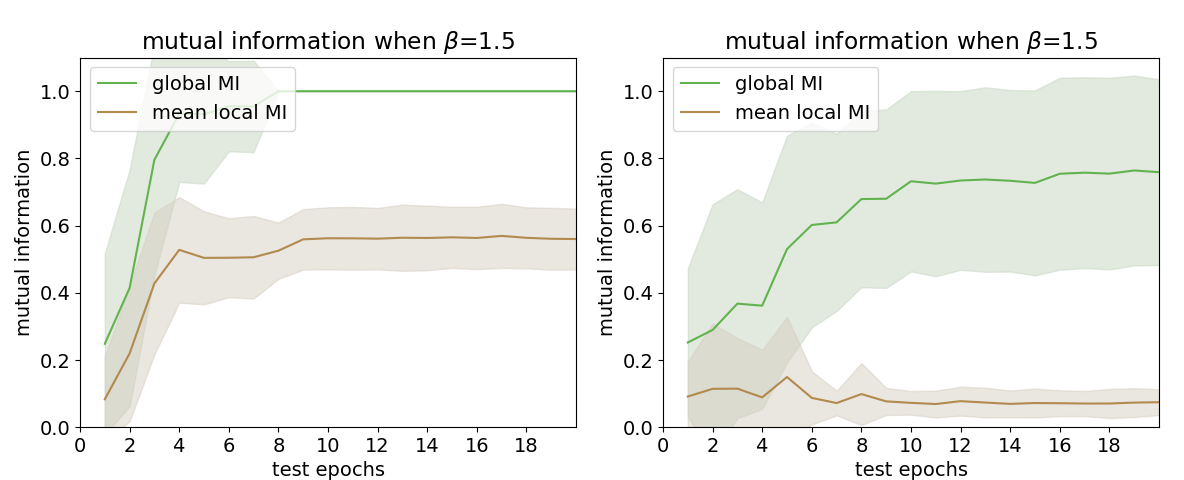}
    \caption{Mutual information curve in toy experiment. Global MI refers to the mutual information related to the global state, and mean local MI refers to the average value of mutual information related to two individual states.}
    \label{toy}
\end{center}
\end{wrapfigure}
Considering a one-step game, two agents receive a one-bit observation $x^{(1
)},x^{(2)}$ respectively, randomly drawn from $\{0,1\}$, and the agents each take a one-bit action $u^{(1)},u^{(2)}$. The successive observation is simply computed by $x'=x\:\texttt{xor}\:u$. Two typical examples of optimal solutions are listed in Table \ref{EXAM}. The two solutions both hit the maximum of $I(\vect{X}';Z)$ however, in solution B, the latent $Z$ only controls the first state $X'^{(1)}$. We call this phenomena "empowerment degeneracy". In contrary, the solution A shows $Z$ determines the coordination pattern of two agents. When $Z=0$, two agents' behaviour always keeps heterogeneous, and when $Z=1$, two agents' behaviour keeps homogeneous, which is actually the skill of \textbf{coordination}. The key point of solution A is minimizing $I(X'^{(i)};Z)$, which inspires our proposed objective \eqref{OBJ}.\par
We test MASD to demonstrate the ability to avoid empowerment degeneracy. Reward is set by $r_{z}=\log q_{\phi}(z|\vect{x}')-\beta\min_{i=1,2}\log q_{\phi_{i}}(z|x'^{(i)})$. Results are presented in Fig. \ref{toy}. Without the second term in reward, the policies consistently fall into non-coordinating solutions like solution B. Set $\beta=1.5$, MASD succeed to reach solution A, except for several imperfect cases.

\subsection{Visualization of learned skills}
\begin{figure}[htbp]
\centering
\subfigure[]{\label{skill_traj}
\begin{minipage}[t]{0.38\linewidth}
\centering
\includegraphics[height=6.3cm]{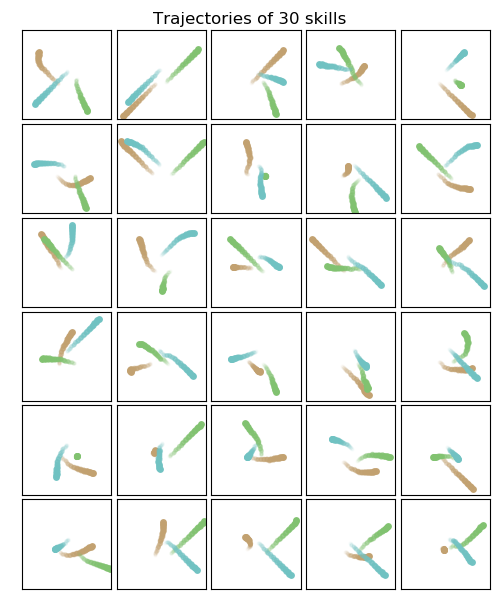}
\end{minipage}%
}%
\subfigure[]{\label{traj_pro_dist}
\begin{minipage}[t]{0.65\linewidth}
\centering
\includegraphics[height=6.3cm]{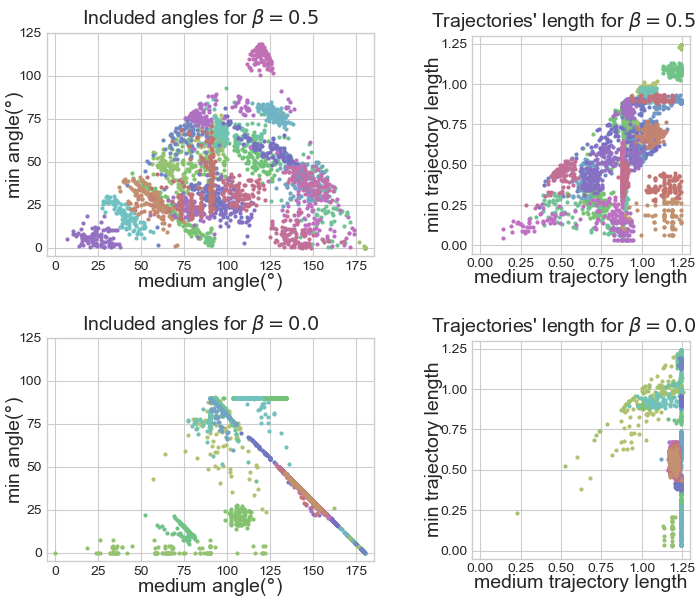}
\end{minipage}%
}%
\centering
\caption{(a) Trajectories of 30 skills from the same initial state. (b) Distributions of trajectories' properties after 100 repeated experiments on slightly disturbed simple spread environment. Left part is the distribution of the two smallest angles and right part is the distribution of the two shortest path length of trajectories. Figures above have $\beta=0.5$ and learned 30 skills, while figures below have $\beta=0.0$ and learned 17 skills.}
\end{figure}

We visualize the learned skills in OpenAI multi-agent particle environments used in \citep{lowe2017multi}. Specifically, we applied our method to the "simple spread" task, in which several agents are rewarded by covering all the landmarks while avoiding collisions. We use up to 30 discrete latent codes to explicitly represent the skill $Z$, and convert the code to one-hot vector. A curriculum approach similar to \citep{achiam2018variational} has been applied to overcome the training difficulty triggered by large latent space. In brief, we start with handful of skills and enlarge the skill set when $\mathbb{E}[\log{q_{\phi}(z|\vect{x})}]$ reaches a high threshold, i.e., $\mathbb{E}[\log{q_{\phi}(z|\vect{x})}]\geq-0.18$. For skill discovery procedure we set the reward of environment to zero and draw the trajectory of all agents after 10000 episodes of training in Fig.~\ref{skill_traj}. To get clear observation, we fix the initial states of the environment when testing. The trajectory patterns of different skills show significant differences even in the environment without reward. 
To verify the diversity of trajectory pattern emerges from the skill diversity rather than randomness of environment or policy, we add random disturbance ($\sigma\sim U[-0.1,+0.1]\times$ world width) to the initial position of the agents and repeat the experiment for 100 times. We calculate some properties of all trajectories in Fig.~\ref{traj_pro_dist}. The left part represents the distribution of the smaller two of the included angles of the three trajectories, while the right part represents the length distribution of the shorter two trajectories. Each skill has its own color. In the two figures above we use coefficient $\beta=0.5$ for $H(Z|f(X^{(i)}))$ and succeed in learning 30 skills, while in the two figures below we set $\beta=0$ for contrast and only 17 skills are learned. When $H(Z|f(X^{(i)}))$ is taken into account, each skill has obvious clustering characteristics in both the dimension of included angle or length of the trajectories.
\subsection{Local entropy of learned skills}
\begin{figure}[htbp]
    \centering
    \subfigure[]{\label{error}
        \includegraphics[height=2.7cm]{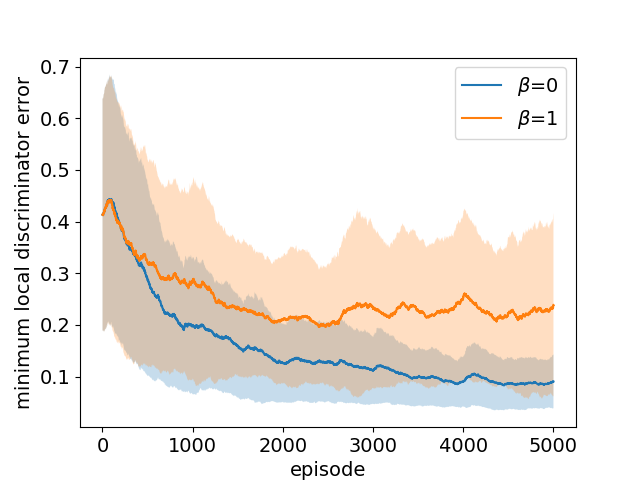}
    }
    \subfigure[]{\label{demo}
        \includegraphics[height=2.7cm]{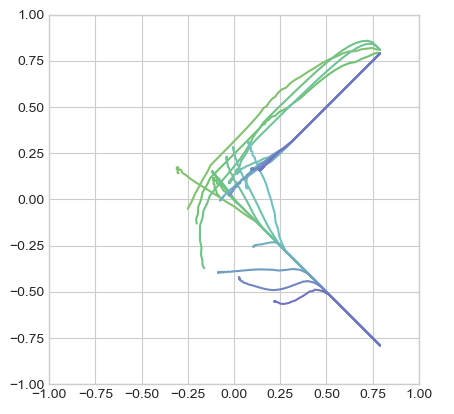}
    }
    \subfigure[]{\label{beta0}
        \includegraphics[height=2.7cm]{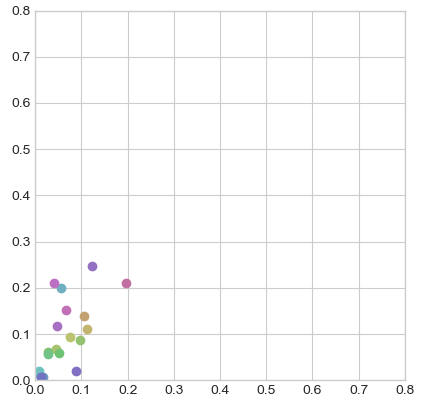}
    }
    \subfigure[]{\label{beta1}
        \includegraphics[height=2.7cm]{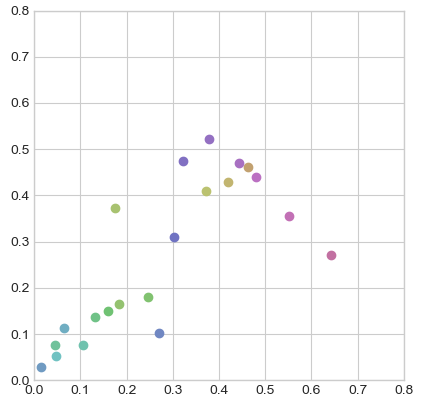}
    }
    \caption{(a) Minimum $L_{1}$ prediction error across all local discriminators. (b) Behaviour of learned policies. (c) Standard deviation of the last position of agent 1, calculated across 16 initial conditions. Each dot represents the standard deviation of one skill. $\beta=0$. (d) The same with (c), $\beta=1$}
    \label{fig4}
\end{figure}
MASD aims to augment the opaqueness of $Z$ from local observations. It will eventually raise the local entropy of skills, characterized as $H(X^{(i)}|Z)$. We find clues in "rendezvous" environment. In this environment, the agents are trained by pseudo reward, combined with a weak signal $-0.1\max_{i}d_{i}^{2}$ ($d_{i}$ is the distance from the central point). The weal signal encourages all agents moving to the central point. As described in Fig.\ref{fig4}, with $\beta=1$, each skill is more diverse on the level of local state, which indirectly confirms that MASD skills focus more on coordination rather than individual patterns.
\subsection{Learning with pretrained models}
To examine the role of skills learned by MASD in specific task, we apply our method to the "simple tag" task, a classical predator-and-prey multi-agent environment that is more complex than "simple spread". We find that models pretrained with MASD has better performance than models with random initialization on performance. 
\begin{figure}[htbp]
\centering
\subfigure[]{\label{chase}
\begin{minipage}[t]{0.5\linewidth}
\centering
\includegraphics[height=5.3cm]{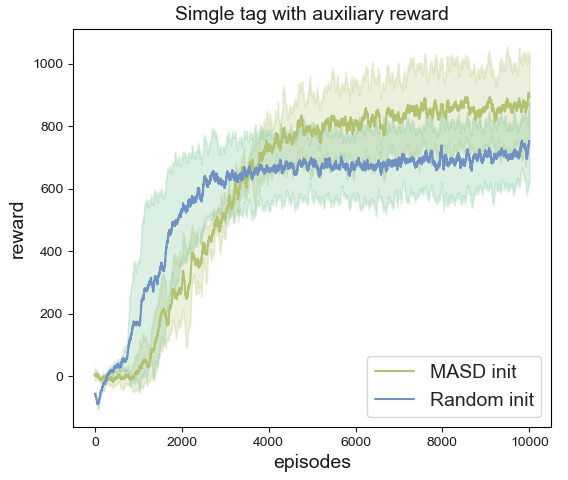}
\end{minipage}%
}%
\subfigure[]{\label{chaseb}
\begin{minipage}[t]{0.5\linewidth}
\centering
\includegraphics[height=5.3cm]{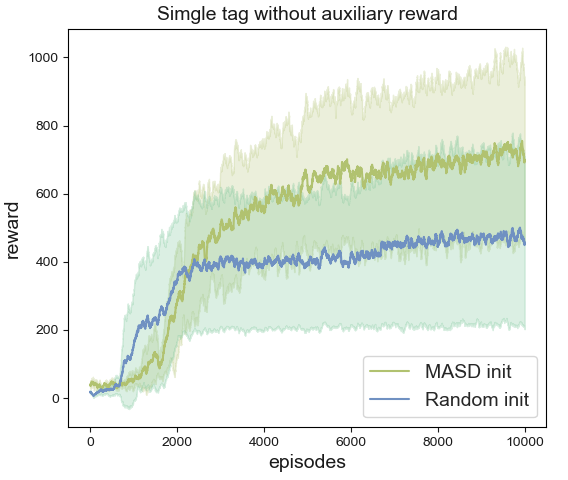}
\end{minipage}%
}
\caption{Reward curves of different initialization. We fix the skill of MASD initialization during training by choosing the skill with the highest reward at the start of training.}
\end{figure}

Specifically, the goal of the agents is to cooperate in the pursuit of a randomly moving prey whose speed is higher. There are two parts of the reward: one is the goal reached reward when one agent hit the prey, and another is the auxiliary reward related to the distance between agent and prey. We use MADDPG to train a randomly initialized model and a model initialized with MASD separately. The reward curves of 5 seeds are plotted in Fig. \ref{chase}. The models initialized with MASD have convergence reward 150 higher on average. When we remove the auxiliary reward to make the task more difficult, the reward curves of 5 seeds are plotted in Fig. \ref{chaseb}. Models initialized with MASD get reward 700 on average, while models randomly initialized only get reward 450 on average. The results suggest MASD pretrained model may gain advantage of performance through skill learning.

\section{Related work}
Reinforcement learning as graphical-model probabilistic inference has been studied in prior works \citep{10.5555/2049078, ziebart2008maximum, furmston2010variational, levine2018reinforcement}. This framework leads to an augmented objective to maximize entropy which provides an alternative way to encourage exploration \citep{haarnoja2017reinforcement, haarnoja2018soft, liu2017stein}. In recent papers, latent space is introduced to model the latent structure of agent policy explicitly \citep{houthooft2016vime, pmlr-v80-igl18a, haarnoja2018latent, hausman2018learning}. Since mutual information can be perceived as the measure of empowerment \citep{mohamed2015variational}, by maximizing mutual information, the agent can learn a set of diverse skills while the skills encoded as latent variable are easy to infer from states or trajectories \citep{gregor2016variational, achiam2018variational, sharma2019dynamics}. DIAYN demonstrates the skills learned without task-specific reward provide a good initialization to successive learning, serve as options in hierarchical reinforcement learning, or imitate en expert \citep{eysenbach2018diversity}. Regarding multi-agent reinforcement learning, \citet{mahajan2019maven} adopts a latent policy to implement committed exploration in multi-agent Q-learning algorithms. However, to our best knowledge, our approach is the first on unsupervised skill discovery of coordination in MARL.\par
Coordination of agents is crucial to MARL, especially in cooperative settings requiring agents to reach a collective goal \citep{cao2012overview}. Part of methods concentrate on the credit/role assignment problem to decompose the collective reward to each agent \citep{pmlr-v80-rashid18a, foerster2018counterfactual, le2017coordinated}. Other works focus on the mechanism of information exchange, i.e., learning communication protocols between the agents \citep{sukhbaatar2016learning}. Instead of dedicated differentiable communication channel, \citet{lowe2017multi} has proposed MADDPG using centralized Q-functions that take all actions as input. However, the coordination patterns largely depend on the nature of the task goal when reaching the goal needs coordination. Since our paper is coping with unsupervised multi-agent environments, the collective optimization objective should be astutely designed to incentivize autonomous emergence of coordination.\par
Our method can be interpreted as an "information bottleneck" between the latent variable and the global state. In previous work, information bottleneck is a technique for regularizing \citep{tishby2015deep, peng2018variational}. Generally speaking, the bottleneck improves generalization and pushes the intermediate representation being irrelevant to input. Similar to this idea, DIAYN sets a bottleneck between $Z$ and $S$ by minimizing $I(A;Z|S)$. This technique results in a maximum entropy policy \citep{eysenbach2018diversity}. From theoretical analysis and empirical results, the bottleneck in MASD also leads to more diverse policies with higher entropy.
\section{Discussion}
In this work, we have developed MASD, an algorithm allows multiple agents to learn various coordination skills without task-specific reward. We show empowerment degeneracy when maximizing the mutual information between latent variable and the global state. To obtain skills on the level of coordination, we add a regularizer to increase the opaqueness of $Z$ in individual states $f(X^{(i)})$. Empirically we demonstrate our method successfully overcomes empowerment degeneracy while keeps different skills discriminable. \par
Reduction of mutual information between $Z$ and individual agent states brings adversarial term in our objective. As discussed in prior work \citep{pmlr-v70-arjovsky17a}, adversarial objective incurs difficulties in training, also in our experiments, adversarial training significantly slows down the skill learning process. To eliminate adversarial learning, we can investigate prior-involved method to autonomously learn coordination patterns, which means we need a good representation of the relation of multiple agents. This research is left as future work.

{\small
\bibliographystyle{plainnat}
\bibliography{refer}
}

\end{document}